\documentclass[10pt, twocolumn]{IEEEtran}
\IEEEoverridecommandlockouts
\usepackage{amsmath, graphics,amssymb,epsfig,subfigure,amsthm,mathtools}
\usepackage{cite}
\usepackage{color}
\usepackage{url}
\usepackage{color,soul}

\newtheorem{theorem}{Theorem}
\newtheorem{lemma}{Lemma}

\newtheorem{corollary}{Corollary}

\newcommand{\be}{\begin{equation}}

\newcommand{\ee}{\end{equation}}
\newcommand{\bea}{\begin{eqnarray}}
\newcommand{\eea}{\end{eqnarray}}
\newcommand{\bdp}{\begin{displaymath}}
\newcommand{\edp}{\end{displaymath}}
\usepackage{graphicx,amsmath}

\makeatletter
\def\hlinewd#1{%
\noalign{\ifnum0=`}\fi\hrule \@height #1 \futurelet \reserved@a\@xhline}

\newcommand{\RN}[1]{%
  \textup{\uppercase\expandafter{\romannumeral#1}}%
}

\makeatother

\begin{document}

\title{\huge{Online Power Allocation at Energy Harvesting Transmitter for Multiple Receivers with and without Individual Rate Constraints for OMA and NOMA Transmissions}}
\author{
\IEEEauthorblockN{Mateen Ashraf, Zijian Wang and Luc Vandendorpe, \textit{Fellow,  IEEE}}\\
\IEEEauthorblockN{
 Universit\'e catholique de Louvain, Louvain-la-Neuve, Belgium \\
Email: \{mateen.ashraf, zijian.wang, luc.vandendorpe\}@uclouvain.be
}
} \maketitle \thispagestyle{empty}

\begin{abstract}
In this paper, we propose an online power allocation scheme to maximize the time averaged sum rate for multiple downlink receivers with energy harvesting transmitter. The transmitter employs non-orthogonal multiple access (NOMA) and/or orthogonal multiple access (OMA) to transmit data to multiple users. Additionally, we consider the scenario where each individual user has a quality of service constraint on its required instantaneous rate. The decisions of total transmit power and power allocation for different users in a given time slot are obtained with the help of Lyapunov optimization technique. The proposed schemes do not require any statistical information of the channel states and the harvested energy. The proposed power allocation schemes entail in small complexity based power allocation decisions. Therefore, the proposed schemes can provide solutions in real time and are more suited for online power allocation problems where the system state parameters (e.g. channel state, harvested energy etc) change quickly. The performances of the proposed schemes are demonstrated with the help of simulation results.

\end{abstract}
\section{Introduction} \label{sec:intro}

The role of energy harvesting technique for improving the lifetime of the wireless networks has been recognized recently [1]. In an energy harvesting wireless network, the energy can be harvested either from radio frequency transmissions or from renewable energy sources. The most practical examples of energy harvesting from renewable sources include energy harvesting from solar radiations, vibrations and wind. Among other performance metrics, an important factor that has been widely used in the literature to assess the performance of the wireless networks comprising of multiple users is the sum rate. Since the transmission rate is a function of allocated resources, a possible way to improve the sum rate is to efficiently allocate the power/bandwidth among different users so that the overall sum rate is maximized. Nevertheless, this becomes a difficult task when the sum rate is to be maximized over a time horizon.

A few off-line power allocation strategies were
proposed for multiple access
systems [2] and fading channels [3]. In their considered system models, the authors of both [2] and [3] assumed that harvested energy is known apriori.
However, in a practical setting, such a non-causal assumption is not valid. To address this issue, several power allocation strategies were proposed in the literature
to maximize the transmission rate [4], [5]. However, their proposed solutions assume the availability of the information related to the statistics of the harvested energy and channel fading at the transmitter. In addition, their proposed solutions entail high complexity algorithms which limit their application to online power allocation problems. Therefore, a low complexity online sum rate maximizing power allocation scheme is required that can properly function only on the basis of causal system state information.

There are several existing power allocation algorithms that operate without requiring the statistical knowledge of the system state information. For instance, the dual stochastic optimization techniques have been widely utilized in the literature to optimize transmission rate dependent utility functions. In [6], the ergodic transmission rate was maximized by using a stochastic descent-based algorithm. A cross-layer resource allocation scheme was proposed in [7] to optimize linear and logarithmic functions of the throughput. 
A weighted sum rate maximization algorithm was proposed in [8] while considering energy causality and data queue stability constraints. A multi-input multi-output
downlink system with energy trading between a base station
(BS) and the main grid was investigated in [9] for maximizing the throughput while satisfying the constraint on the cost of energy.

In recent times, Amirnavaei \textit{et al} [10] proposed an online power allocation method for time average sum rate maximization. Their proposed solution was based on the Lyapunov optimization technique and it did not require the statistical information on the random processes. The same authors investigated a cooperative communication system [11] and proposed an online power allocation scheme for maximizing the time-averaged sum rate. However, a single user scenario was considered in these works. More recently, an energy efficiency maximization problem was considered in [12] for only two users case. In their work, the total bandwidth and power is equally split between the two users and therefore the possibility of inter-user interference was not considered. The equal division of power and bandwidth between the users may be a good solution when the channel gains of different users are identical however such an equal distribution of resources may lead to performance degradation in scenarios with a greater variation among channel gains of different users. To the best of our knowledge, none of the existing works has considered multiuser interference scenario. In this work, we consider a downlink system model with $K\geq 2$ users where inter-user interference exists among the users.  The considered scenario in this paper is not only much more complex than [12] but also more realistic.

For facilitating $K$ multiple users in the downlink, we assume that the transmitter uses non-orthogonal multiple access (NOMA) and orthogonal multiple access (OMA) scheme for downlink transmission. The single user problem has already been studied in the literature [10] and our considered system model encompasses that existing work by putting $K=1$. Note that NOMA has been used extensively for downlink systems with more than one receiver [13]-[15]. It is widely known that the sum rate performance of the NOMA is better than that of the OMA if the channel gains for different receivers have greater variation. Owing to this reason, the NOMA is considered to be a promising technology in the 5G wireless communication standards. On the other hand, NOMA with successive interference cancellation results in increased complexity at the receivers. This is owing to the fact that the stronger users in NOMA scheme have to decode the information of the weaker users and subsequently subtract it from the received signal to perform inter-user interference cancellation. This operation causes the increase in the complexity of the receivers, especially if the number of downlink users are more than $2$. In such a situation, it may be more suitable to implement OMA for multiple downlink users. 

Additionally, in a practical scenario it is reasonable to assume that the individual downlink users have minimum data rate requirements that should be met for all the time slots. Under this scenario, the problem not only becomes more complex but it may also be infeasible. The infeasibility may arise from the random nature of the channel gains as well as from the random nature of the harvested energy. For example, it is possible that the channel gains become so low in a particular time slot that the rate constraints cannot be met even by transmitting the whole energy of the battery during a particular time slot. To tackle this issue, we assume that the channel gains remain higher than certain threshold and the harvested energy in each time slot is also higher than a certain threshold.   

In this paper, we relax the battery dynamic constraints as a time-averaged expression. Next, we adopt Lyapunov optimization framework and introduce a virtual queue for the battery. With the relaxed constraint and the introduced virtual queue, we reformulate the time-averaged sum rate maximization problem as a queue stability problem and obtain a suboptimal online power allocation for time-averaged sum rate maximization by solving the queue stability problem. In summary, our aim is to devise a suboptimal solution that satisfies the constraints of the original optimization problem and subsequently show that the objective value achieved by the suboptimal solution has a bounded gap with the actual optimal value of the original optimization problem.

The main contributions of this paper can be outlined as follows:
\begin{itemize}
\item We propose an online power allocation scheme at the BS for long-term average sum rate maximization for multiple downlink users when the BS uses NOMA for downlink transmission in the absence of individual users rate requirements. This scenario is referred to as NOMA-WoR.
\item Then, the proposed power allocation scheme for NOMA case is extended to cover the possibility of individual minimum rate requirements. This scenario is referred to as NOMA-WR. 
\item We propose an online power allocation scheme at the BS for long-term average sum rate maximization for multiple downlink users when the BS employs OMA for downlink transmission in the absence of individual users rate requirements. This scenario is referred to as OMA-WoR.
\item The proposed power allocation scheme for OMA case is also extended to include the possibility of individual minimum rate requirements. This scenario is referred to as OMA-WR.
\item Theoretically, it is shown that the proposed power allocation schemes have bounded performance gap when compared with their optimal power allocation schemes.
\item Simulation results are presented to demonstrate the effectiveness of the proposed schemes.
\end{itemize}

The rest of the paper is organized as follows. Section II presents the system model and the related assumptions. The online power allocation schemes for the NOMA case with and without individual rate requirements are presented in Section III. In Section IV, we present the online power allocation schemes for the OMA case with and without individual rate requirements. The performance analysis is provided in Section V. Section VI presents the simulation results. Finally, the paper is concluded in Section VII.

\section{System Model and Problem Formulation} \label{sec:model}


The system model comprises of a transmitter that is equipped with the energy harvesting device and supports $K \geq 2$ receivers. A pictorial representation of the considered system is provided in Fig. 1 and the important system parameters are outlined in Table I.
\begin{figure}[h]
\begin{center}
\includegraphics[width=3.5in]{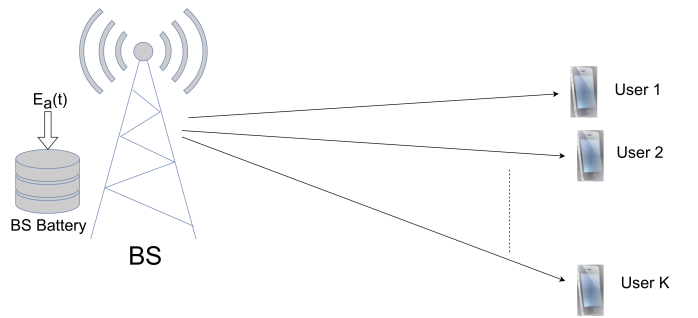}
\end{center}
\caption{System model with energy harvesting base station and multiple downlink users.}
\end{figure}
\begin{table}
\caption {Important System Parameters}
\begin{center}
\begin{tabular}{|c|c|}
		\hline
		Parameter & Notation  \\
		\hline
		battery status at time slot $t$ & $E_b(t)$ \\
		\hline
		harvested Energy at time slot $t$ & $E_h(t)$ \\
		\hline
		minimum battery level & $E_{min}$\\
		\hline
		maximum battery level & $E_{max}$ \\
		\hline
		maximum chargeable energy & $E_{c, max}$\\
		\hline
		arrived energy at time slot $t$ & $E_a(t)$ \\
		\hline
		maximum harvestable energy & $E_{h,max}$\\
		\hline
		time slot duration & $\Delta t$ \\
		\hline
		$k$-th user channel gain during time slot $t$ & $\gamma_{k}(t)$\\
		\hline
		$k$-th user power allocation ratio during time slot $t$ & $\rho_k(t)$ \\
		\hline
		$k$-th user bandwidth allocation ratio during time slot $t$ & $\alpha_k(t)$\\
		\hline
		power allocation ratio vector & $\boldsymbol{\rho}(t)$\\
		\hline
		bandwidth allocation ratio vector & $\boldsymbol{\alpha}(t)$\\
		\hline
		minimum channel gain & $\gamma_{min}$\\
		\hline
		maximum channel gain & $\gamma_{max}$ \\
		\hline
		transmit power during time slot $t$ & $P(t)$\\
		\hline
		maximum transmit power & $P_{max}$ \\
		\hline
		minimum required power to satisfy the rate constraints & \textcolor{white}{0} \\ when all the channels have 					strongest possible values & $P_{th}^{best}$  \\
		\hline
		minimum required power to satisfy the rate constraints & \textcolor{white}{0} \\ when all the channels have worst possible values & $P_{th}^{worst}$  \\
		\hline
	\end{tabular}
	\end{center}
	\end{table}
	
We assume that the total time is divided into $T$ time slots where each time slot is of duration $\Delta t$. The individual time slots are indexed by $t$. The channel gain $|h_k(t)|^2$ between the transmitter and the $k$-th receiver $(k\in \{1,2\cdots K\})$ remains constant during the time $t$. This means that the channels are block fading and their value remains constant during each time slot while it can vary across different time slots. The additive white Gaussian noise at the $k$-th receiver is denoted by $n_k(t)\sim\mathcal{CN}(0,\sigma^2)$.
Let us denote $E_h(t)$ as the amount of the harvested energy into the battery at the transmitter at each time slot $t$. The battery level at time slot $t$ is represented by $E_b(t)$. We assume that $E_{\min} \leq E_b(t) \leq E_{\max}$, where $E_{\min},E_{\max}$ are the lower and upper bounds on the energy levels in the battery, respectively.

Let $E_{c,\max}$ and $P_{\max}$ be the maximum charging amount and the maximum transmit power which satisfies ${\Delta t P_{\max} \leq E_{\max}-E_{\min}}$. We assume that $E_{c,\max} \leq \Delta t P_{\max}$. Let us define $P(t)$ as the transmit power at each time slot $t$, which remains constant over each individual time slot. The transmit power is constrained as $0\leq P(t) \leq P_{\max}, \forall t$.

The evolution of $E_b(t)$ over time can be written as [16] 
\bea
E_b(t+1) = E_b(t) -{\Delta}tP(t)+ E_h(t).\label{dynamics}
\eea
Then, $P(t)$ is bounded by ${\Delta t P(t)\leq E_b(t)-E_{\min}}$ for all $t$.
The harvested energy $E_h(t)$ has following constraint [11]
\bea
0 \leq E_{h}(t)\leq \mathrm{min}\{E_{c,max},E_a(t)\},\nonumber
\eea
where $E_a(t)$ denotes the arrived energy. From the battery capacity constraint and dynamics of battery, we can write
\bea
0 \leq E_h(t)\leq \mathrm{min}\{E_{c,max},E_a(t),E_{h,max}(t)\},\label{E_h(t)}
\eea
where $E_{h,max}(t)$ in (2) represents the maximum harvestable energy which is defined as $E_{h,max}(t)=E_{max}-E_{b}(t)+{\Delta}t P(t)$. The constraint on the harvested energy actually means that the harvested energy $E_h(t)$ should be such that the battery level $E_b(t)\leq E_{max}$ for all $t$.

With NOMA, assuming that the $k$-th user is allocated with $\rho_k(t)$ portion of the total transmit power $P(t)$ during time slot $t$, we can write the rate of the $k$-th user during time slot $t$ as
\bea
R_k^{NOMA}(t)=\log\left(1+\frac{\rho_k(t)P(t)\gamma_k(t)}{P(t)\gamma_k(t)\sum_{i=k+1}^K\rho_i(t)+1}\right),
\eea
where $\gamma_k(t)\triangleq |h_k(t)|^2/\sigma^2$ with $\gamma_k(t)< \gamma_{k+1}(t)< \gamma_{max}$ for $k\in\{1,\cdots ,K-1\}$. Here, we assume that the downlink users employ successive interference cancellation to decode their desired signal and therefore the interference due to the weaker users is removed in (3). In this regard, we impose a power order constraint as $\rho_1(t)\geq \rho_2(t)\cdots \geq \rho_K(t)$. It can be easily verified that in the absence of the power order constraint, the power allocation problem to maximize the sum rate for multiple users translate into the rate maximization for single user with the highest channel gain\footnote{See the result presented in Lemma 5 below.}. By introducing the new variables $z_k(t)=\sum_{i=k}^K\rho_i(t)$, the power order constraint is converted to following constraint
\bea
z_1(t)-z_2(t)\geq z_2(t)-z_3(t)\geq \cdots \geq z_K(t)\geq 0.
\eea
The sum rate can be written as 
\bea
\sum_{k=1}^KR_k^{NOMA}(t)=\sum_{k=1}^KG_k(z_k(t),P(t)),
\eea
where $G_k(z_k(t),P(t))=\log\left(1+\gamma_k(t)P(t)z_k(t)\right)-\log\left(1+\gamma_{k-1}(t)P(t)z_k(t)\right)$ for $k\in\{2,3\cdots K\}$ and $G_1(z_1(t),P(t))=\log\left(1+\gamma_1(t)P(t)z_1(t)\right)$.
Then, we propose a power allocation scheme to maximize the time-averaged sum rate over all the users while guaranteeing the operational constraints of the battery.

The optimization problem can be written as
\begin{align}
\textbf{P1}
\max_{P(t), \mathbf{z}(t)}&\frac{1}{T}\lim_{T\rightarrow \infty}\sum_{t=0}^{T-1}\mathbb{E}\left(\sum_{k=1}^KG_k(z_k(t),P(t))\right)\label{P1}\\
\text{s.t. }&0\leq P(t) \leq P_{\max},~~ \tag{6a}\nonumber\\
& {\Delta t P(t)\leq E_b(t)-E_{\min}}, \tag{6b}\nonumber\\
&(1),\tag{6c}\nonumber\\
&z_1(t)\leq 1,~~\tag{6d} \nonumber\\
&(4), \tag{6e}\nonumber \\
&E_h(t)\leq \mathrm{min}\{E_{c,max},E_a(t),E_{h,max}(t)\}, \tag{6f}  
\end{align}
where $\mathbf{z}(t)=\{z_1(t),\cdots,z_K(t)\}$ and $\mathbb{E}(.)$ represents the expectation with respect to the system state.

For the case of OMA, we assume that the total bandwidth\footnote{For simplicity of exposition, we assume $W=1$ however the subsequent discussion is also applicable to any general value of $W$.} $W$ is orthogonally distributed among multiple users. The bandwidth allocated to the $k$-th user during $t$-th time slot is denoted by $\alpha_k(t)$. Then, the information rate for the $k$-th user can be written as 
\bea
R_k^{OMA}(t)=\alpha_k(t)\log\left(1+\frac{\rho_k(t)P(t)\gamma_k(t)}{\alpha_k(t)}\right).
\eea
In addition to the existing constraints in problem P1, we impose following constraints on the individual bandwidth allocations
\bea
0\leq \alpha_k(t)\leq 1,\\
0\leq \sum_{k}\alpha_k(t)\leq 1.
\eea
With these constrains added, the long-term sum rate maximization problem for OMA case can be written as follows
\begin{align}
\textbf{P2}
\max_{\{P(t), \boldsymbol{\rho}(t),\boldsymbol{\alpha}(t)\}}&\frac{1}{T}\lim_{T\rightarrow \infty}\sum_{t=0}^{T-1}\mathbb{E}\left(\sum_{k=1}^KR_k^{OMA}(t)\right)\label{P1}\\
\text{s.t. }&\text{6a, 6b, 6f} ~~\tag{10a}\nonumber \\
&0\leq \sum_{k=1}^{K}\rho_k(t)\leq 1,~~ \tag{10b}\nonumber\\
&0\leq \rho_k(t) \leq 1,~~ \tag{10c} \nonumber \\
& 0 \leq \alpha_k(t)\leq 1, ~~ \tag{10d}\nonumber \\
& \sum_{k=1}^K\alpha_k(t)\leq 1.~~ \tag{10e} \nonumber
\end{align}

Due to the randomness of the harvested energy and the channel gains, the optimization problems \textbf{P1} and \textbf{P2} are difficult to solve. In addition, the constraints depend on $E_b(t)$, which has time-coupling dynamics over time.

This results in the power allocation decisions $\{P(t)\}$ being correlated over time.
Although dynamic programming (DP) can be used to solve problems \textbf{P1} and \textbf{P2} when the random processes $\{\gamma_k(t)\}$ and $\{E_h(t)\}$ are Markov and their statistics are known, this approach typically results in a high computational complexity. Thus, it is not easy to provide a solution efficiently. Moreover, in a real scenario, the statistical knowledge of $\{\gamma_k(t)\}$ and $\{E_h(t)\}$ may not be available in advance, which makes such an assumption less practical.

In this paper, our aim is to devise a low complexity online sum rate maximizing power allocation scheme which does not require the statistical information of $\{\gamma_k(t)\}$ and $\{E_h(t)\}$.
To achieve this goal, we resort to the Lyapunov optimization framework [17] and transform time-average optimization problem into a queue stability problem.
As a first step in this direction and in order to apply the Lyapunov optimization technique, we relax the time-coupled dynamics on the time slot constraints $E_b(t)$, $E_h(t)$, and $P(t)$ by adopting the time-average relation.

From (\ref{dynamics}), we have the following relation of the battery dynamics over time $T$
\bea
E_b(T)-E_b(0) = \sum_{t=0}^{T-1}(E_h(t)-\Delta t P(t)).\label{relation}
\eea
Then, after some mathematical manipulations with $T\rightarrow \infty$, we can represent the time-average relationship between $\bar{E}_h$ and $\bar{P}$ as
\bea
\bar{E}_h-\Delta t \bar{P} =0,\label{long_term}
\eea
where $\bar{E}_h=\lim_{T\rightarrow\infty}\frac{1}{T}\sum_{t=0}^{T-1}\mathbb{E}(E_h(t))$ and $\bar{P}=\lim_{T\rightarrow\infty}\frac{1}{T}\sum_{t=0}^{T-1}\mathbb{E}(P(t))$ [16].

\section{Online Rate Maximizing Power Allocation for Non-Orthogonal Multiple Access} \label{sec:design}
In this section, we solve the online power allocation problem when downlink transmission employs NOMA for multiple downlink users. First, we present the power allocation design without individual user rate constraints. Then, we present the power allocation design with individual rate constraints.
\subsection{Non-Orthogonal Multiple Access without Individual User Rate Constraints (NOMA-WoR)}
The optimization problem in \textbf{P1} can be relaxed by replacing the dynamics in (1) with the long-term time-average constraint (\ref{long_term}) and by removing (6b). After doing these steps, we have the relaxed optimization problem of \textbf{P1} as follows
\begin{align}
\textbf{P3}
\max_{P(t), \mathbf{z}(t)}  &\lim_{T\rightarrow \infty}\frac{1}{T}\sum_{t=0}^{T-1}\mathbb{E}\left(\sum_{k=1}^KG_k(z_k(t),P(t))\right)\label{P2}\\
\text{s.t.}~~~~ &\bar{E}_h-\Delta t \bar{P}=0, \tag{13a}\nonumber\\
&\text{6a, 6d, 6e}\nonumber,\tag{13b}\\
&E_{h}(t)\leq \mathrm{min}\{E_{c,max},E_a(t)\}.\tag{13c}\nonumber
\end{align}
The use of (12) instead of (6b) has allowed us to replace (6f) with (13c) [11]. The constraint (13c) is independent of $P(t)$ however we include it for the sake of completeness [11]. Still, there are two major challenges imposed by \textbf{P3}: First, allocation of power in different time slots such that constraint (13a) is met. Second, the unavailability of the statistical information on the system state makes it difficult to solve \textbf{P3}. In the rest of this section, we devise a suboptimal solution that does not require statistical information of the system state and uses Lyapunov optimization theory to guarantee that (13a) is met. 

To this end, we introduce the virtual queue $Q(t)$ for $E_b(t)$ as
$Q(t) = E_b(t) - C$,\texttt{}
for some constant $C$. The value of $C$ will be obtained later in this section.
Note that stabilizing the queue $Q(t)$ is equivalent to satisfying (13a) [10].
The evolution of the virtual queue with respect to time can be written as 
\bea
Q(t+1) = Q(t) + E_h(t) -\Delta t P(t). \label{X(t)}
\eea
For time slot $t$, the drift-plus-cost metric with the constant $V>0$ is given by
\bea
\mathcal{D} = \Delta(Q(t))-V \mathbb{E}\left(\sum_{k=1}^KG_k(z_k(t),P(t))|Q(t)\right),\label{DPCM}\nonumber
\eea

where $\Delta (Q(t))$ is given as 
\bea
\Delta (Q(t))=\mathbb{E}\left[\frac{Q^2(t+1)}{2}-\frac{Q^2(t)}{2}\bigg|Q(t)\right],
\eea
where $\mathbb{E}[.|Q(t)]$ denotes the expectation with respect to system state given $Q(t)$. The drift-plus-cost metric is a weighted sum of the per-slot Lyapunov drift and the objective function conditioned on $Q(t)$. It is well known that minimizing the drift-plus-cost metric results in stabilizing the virtual queue while optimizing the objective function [10]. For any value of $Q(t)$ and $V\geq0$, it can be easily shown that the upper bound on drift-plus-cost metric $\mathcal{D}$ is [10]
\bea
\mathcal{D}\leq \phi + Q(t)\mathbb{E}\left(E_h(t)-\Delta t P(t)|Q(t)\right) \nonumber \\ -V\mathbb{E}\left(\sum_{k=1}^KG_k(z_k(t),P(t))|Q(t)\right),\label{DPCM_UB}
\eea
where $\phi = {\Delta}t^{2}P_{\max}^2/2$. We will see in Section V that the mathematical form of the upper bound on the drift-plus-cost metric is useful for comparing the objective value achieved by the proposed scheme with the objective value achieved by solving \textbf{P1}. As we have assumed that the statistical information of the system state is not available, evaluating the right hand side (RHS) of (16) and further minimization is not possible. To address this issue, we aim at minimizing the RHS of (16) on per time slot basis by removing the expectation. This approach has been widely used in the Lyapunov optimization theory [10]-[11], [16], [18]-[20]. Although we ignore the time coupling constraints in the following optimization problem, we will show in section III-C that the solution of the proposed scheme satisfies the time coupling and battery constraints of \textbf{P1}. Hence, we have the following optimization problem
\bea
\textbf{P4}
\min_{P(t), \mathbf{z}(t)} \phi+Q(t)[E_h(t)-\Delta t P(t)]\nonumber~~~~~~~~~~~~~~\\ -V\sum_{k=1}^KG_k(z_k(t),P(t)),\label{P3}\\
\text{s.t.~~~~} \text{9b}.~~~~~~~~~~~~~~~~~~~~~~~~~~~~\nonumber
\eea
Problem \textbf{P4} can be decomposed into two optimization problems. First, we fix $P(t)$ in the interval $[0,P_{max}]$ and minimize with respect to $z_k(t)$'s $\forall k \in \{1,\cdots K\}$ and then minimize with respect to $P(t)$. The optimal objective value achieved by the solution obtained through this decomposition method converges (is actually equal) to that achieved by solving \textbf{P4} through any other method [21, pp. 133].
\subsubsection{Minimization With Respect To $z_{k}(t)$'s}
For a fixed $P(t)$, the optimization problem \textbf{P4} is equivalent to the following problem 
\bea
\textbf{P5}~~~~
\max_{\mathbf{z}(t)} \sum_{k=1}^{K} G_k(z_k(t),P(t)) \label{P5}\\
\text{s.t.~~} \text{6d, 6e}~~~~~~~~~~ \nonumber
\eea
The optimal values of $z_k(t)$ for problem \textbf{P5} are given by the following Lemma.
\begin{lemma}
The solution of \textbf{P5} is given as follows
\bea
z_k^*(t)=\frac{K-k+1}{K}, \forall k \in\{1,2\cdots K\}.
\eea
\end{lemma}
\begin{proof}
First, we note that $z_1(t)=1$ since $\sum_{k=1}^{K} G_k(z_k(t),P(t))$ is increasing in $z_1(t)$. Furthermore, the derivative of $G_k(z_k(t),P(t))$ with respect to $z_k(t)$ is given as 
\bea
\frac{P(t)(\gamma_k(t)-\gamma_{k-1}(t))}{(1+\gamma_k(t)P(t)z_k(t))(1+\gamma_{k-1}(t)P(t)z_k(t))}.
\eea
Since $\gamma_k(t)>\gamma_{k-1}(t)$, we have $\frac{\partial G_k(z_k(t),P(t))}{\partial z_k(t)}\geq 0, \forall k\in\{1,2\cdots K\}$ . This means the constraint $z_K^*(t)\leq z_{K-1}^*(t)-z_K^*(t)$ should be active. This implies $z_{K-1}^*(t)=2z_{K}^*(t)$. In a similar way, we have $z_{K-2}^*=3z_{K}^*(t),\cdots, z_{1}^*(t)=Kz_{K}^*(t)$. Now using the fact that $z_1^*(t)=1$, one can easily see that $z_k^*(t)=\frac{K-k+1}{K}$. This completes the proof.
\end{proof}
With the help of Lemma 1 and using the relationships between $\rho_k(t)$'s and $z_k(t)$'s we can see that $\rho_1^*(t)=\rho_2^*(t)=\cdots \rho_K^*(t)=\frac{1}{K}$. This power allocation can also be explained intuitively. On the one hand, we should maximize the power allocated to the strongest user to maximize the sum rate. On the other hand, the power order constraint requires that the power allocated to the weakest users should be largest. This phenomenon results in a power allocation in which each user is allocated equal power.
\subsubsection{Minimization with Respect To $P(t)$}
After putting the optimal power allocations, obtained from Lemma 1, in the objective function of \textbf{P4} and removing the irrelevant terms we get
\bea
\textbf{P6}~~~~
\min_{P(t)} ~~-Q(t)\Delta t P(t)-V\underbrace{\sum_{k=1}^{K}G_k(z_k^*(t),P(t))}_{f_2(P(t))}\label{P4}\\
\text{s.t.~~~~} 0\leq P(t)\leq P_{max}.~~~~~~~~~~~~ \nonumber 
\eea
We have following Lemma for problem \textbf{P6}.
\begin{lemma}
Problem \textbf{P6} is a convex optimization problem.
\end{lemma}
\begin{proof}
Since the first term of the objective function is linear, we only need to show that $\sum_{k=1}^{K}G_k(z_k^*(t),P(t))$ is concave. It is clear from Lemma 1 that $z_{k}^*(t)$ are independent from $P(t)$. This implies that $G_1(z_1^*(t),P(t))$ is concave with respect to $P(t)$. Additionally, the double derivative of $G_k(z_k^*(t),P(t)), \forall k \in \{2,\cdots K\}$ with respect to $P(t)$ is given as
\bea
& \frac{\partial ^2 G_k(z_k^*(t),P(t))}{\partial P^2(t)}=\frac{z_k^{*2}(t)(\gamma_{k-1}^2(t)-\gamma_k^2(t))}{(1+\gamma_{k-1}(t)z_k^*(t)P(t))^2(1+\gamma_k(t)z_k^*(t)P(t))^2}\nonumber \\ &+\frac{2\gamma_{k-1}(t)\gamma_k(t)z_k^{*2}(t)P(t)(\gamma_{k-1}(t)-\gamma_k(t))}{(1+\gamma_{k-1}(t)z_k^*(t)P(t))^2(1+\gamma_k(t)z_k^*(t)P(t))^2}.
\eea
Since $\gamma_k(t)> \gamma_{k-1}(t)$, we have $\frac{\partial ^2 G_k(z_k^*(t),P(t))}{\partial P^2(t)}<0$. Hence, $G_k(z_k^*(t),P(t)), \forall k \in \{1,2,\cdots K\}$ is a concave function. As the non-negative sum of concave functions is a concave function, we conclude that $\sum_{k=1}^{K}G_k(z_k^*(t),P(t))$ is concave with respect to $P(t)$. This completes the proof.
\end{proof}
Although the optimal $P(t)$ does not admit a close form solution, we can use Bisection algorithm to find the solution of \textbf{P6}.

The solution of \textbf{P6}, denoted by $P^{P6}(t)$, in the current form may not result in transmit power decisions that satisfy the constraints on the battery $E_{min}\leq E_b(t)\leq E_{max}$ and the constraint on the transmit power ${\Delta t P^{P6}(t)\leq E_b(t)-E_{\min}}$. Therefore, it is important to find the values of parameters $V$ and $C$ to make sure that these constraints are always met in the proposed power allocation scheme. In the next subsection, we find the values of parameters $V$ and $C$ which make sure that the virtual queue $Q(t)$ remains stable during all the time slots and the constraints $E_{min}\leq E_b(t)\leq E_{max}, {\Delta t P^{P6}(t)\leq E_b(t)-E_{\min}}$ are also satisfied for all $t$. 
\subsubsection{Stabilizing the Virtual Queue}
The non-close form solution of \textbf{P6} poses the problem of finding the appropriate values of $V$ and $C$. However, this difficulty can be addressed with the help of Lemma 3 and Theorem 1 presented below.
\begin{lemma}
The optimal transmit power for the following problem
\bea
\textbf{P7}~~~~
\min_{P(t)}~~ -Q(t)\Delta t P(t)-V\underbrace{\log(1+P(t)\gamma_K(t))}_{f_1(P(t))}\label{P5}
\eea
\bea
\text{s.t.}~~~~ 0\leq P(t)\leq P_{max}.~~~~~~~~~~~~ \nonumber 
\eea
is given as 
\begin{align}
P^{P7}(t)= \begin{cases}
P_{max} &\text{for } Q(t)>Q_1^{th}(t),\\
\frac{-V}{\Delta t Q(t)}-\frac{1}{\gamma_K(t)} &\text{for } Q_2^{th}(t)\leq Q(t)\leq Q_1^{th}(t), \\
0 & \text{for } Q(t)< Q_2^{th}(t),
\end{cases}
\end{align}
where 
\bea
Q_{1}^{th}(t)= \frac{-V}{\Delta t(P_{max}+\frac{1}{\gamma_K(t)})},~~ Q_2^{th}(t)=-\frac{V\gamma_K(t)}{\Delta t}.\nonumber 
\eea
\end{lemma}
\begin{proof}
See proposition 1 of [10].
\end{proof}
\begin{theorem}
The optimal transmit power for problem \textbf{P6} is always less than or equal to the optimal transmit power for the optimization problem \textbf{P7}.
\end{theorem}
\begin{proof}
First, if $Q(t)>0$ then the optimal transmit power for \textbf{P7} is $P_{max}$. This is because both $Q(t)\Delta tP(t)$ and $\log(1+P(t)\gamma_K(t))$ are increasing in $P(t)$ for $Q(t)>0$. Also, since 
\begin{small}
\begin{align}
\frac{\partial G_k(z_k^*(t),P(t))}{\partial P(t)}=\frac{z_k^*(t)\gamma_k(t)}{(1+\gamma_k(t)z_k^*(t)P(t))(1+\gamma_{k-1}(t)z_k^*(t)P(t))}\nonumber \\ -\frac{z_k^*(t)\gamma_{k-1}(t)}{(1+\gamma_k(t)z_k^*(t)P(t))(1+\gamma_{k-1}(t)z_k^*(t)P(t))},
\end{align}
\end{small}
is positive since $\gamma_k(t)\geq \gamma_{k-1}(t)$. As a result, $\sum_{k=1}^{K}G_k(z_k^*(t),P(t))$ is also increasing in $P(t)$. Hence, the optimal transmit power for \textbf{P7} is also $P_{max}$. Therefore, the optimal transmit power for both problems is $P_{max}$ if $Q(t)>0$.

On the other hand, when $Q(t)<0$ then optimal transmit power for both problems can lie in $[0, P_{max}]$. We note that 
\bea
\log(1+P(t)\gamma_{max})\geq \log(1+P(t)\gamma_{K}(t))\nonumber \\ \geq \sum_{k=1}^KR_k^{NOMA}(t),
\eea
where $R_k^{NOMA}(t)$ is provided in (3). The inequality (26) is due to the following facts.
\begin{itemize}
\item F1: According to Lemma 1, for any fixed transmit power $P(t)$, the sum rate $\sum_{k=1}^KR_k(t)$ is maximized when all the power allocations, $\rho_1(t), \rho_2(t),\cdots \rho_K(t)$, are equal.
\item F2: $\log\left(1+\frac{\frac{1}{K}P(t)\gamma_k(t)}{P(t)\gamma_k(t)\sum_{i=k+1}^K\frac{1}{K}+1}\right)$ is increasing in $\gamma_{k}(t)$.
\item F3: $\sum_{k=1}^K\log\left(1+\frac{\frac{1}{K}P(t)\gamma_K(t)}{P(t)\gamma_K(t)\sum_{i=k+1}^K\frac{1}{K}+1}\right)=\log(1+\gamma_K(t)P(t))$. 
\item F4: $\gamma_k(t)\leq \gamma_{max}, \forall k\in \{1,2 \cdots K\}$.
\end{itemize}
Note that the objective functions of both \textbf{P6} and \textbf{P7} are concave while $-Q(t)\Delta t P(t)$ is increasing if $Q(t)<0$. Furthermore, $-V\log(1+P(t)\gamma_K(t))$ and $-V\sum_{k=1}^KG_k(z_k^*(t),P(t))$ are decreasing. Using (26) it can be easily seen that if the optimal transmit power for problem \textbf{P7} is $P^{P7}(t)$ then the optimal transmit power for \textbf{P6}, $P^{P6}(t)$, is always smaller than $ P^{P7}(t)$ because $-V\sum_{k=1}^KG_k(z_k^*(t),P(t))$ decreases at a slower rate than $-V\log(1+P(t)\gamma_K(t))$. This reasoning can be easily understood with the help of Fig. 2.  
\end{proof}
\begin{figure}[h]
\begin{center}
\includegraphics[width=3.5in]{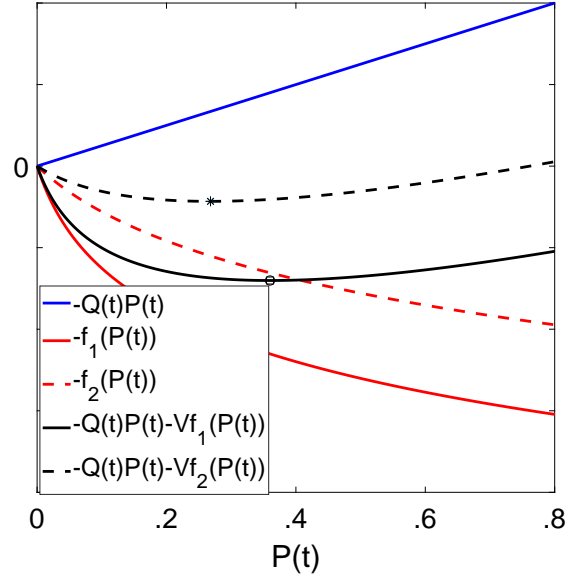}
\end{center}
\caption{Pictorial depiction of the optimal values of transmit power for single user case and multiple user case when $Q(t)<0$ and $\Delta t=1$. The circle and star on black curves represent the optimal values of transmit power for the single user case and multiple user case, respectively.}
\end{figure}
We have the following corollary based on Theorem 1.
\begin{corollary}
The virtual queue $Q(t)$ is bounded as
\bea \label{e5}
-\frac{V\gamma_{max}}{\Delta t}-\Delta t P_{max} \leq Q(t) \leq E_{c,max},
\eea
for all time slots.
\end{corollary}
\begin{proof}
It can be easily verified that if $Q(t)<-\frac{V\gamma_{max}}{\Delta t}$ then optimal transmit power for problem \textbf{P7} is equal to $0$. According to Theorem 1, the optimal transmit power for problem \textbf{P6} will also be zero. This means when $Q(t)<-\frac{V\gamma_{max}}{\Delta t}$, $Q(t+1)$ is always increasing. On the other hand, when $Q(t)\geq -\frac{V\gamma_{max}}{\Delta t}$ we may have $Q(t+1)$ decreasing. However, the maximum decrease occurs when harvested energy is zero and transmit power is $P_{max}$. Hence, it follows
\bea
-\frac{V\gamma_{max}}{\Delta t}-\Delta t P_{max}\leq Q(t).
\eea
The above inequality is true for all the time slots. Therefore, the left hand side (LHS) inequality in (27) is proved.

In a similar way, by checking the possibilities $Q(t)>0$, $Q(t)\leq 0$ and using the fact that $E_{c,max}\leq \Delta t P_{max}$ we can verify that 
\bea
Q(t+1)\leq Q(t)+E_{c,max}\leq E_{c,max},
\eea  
is valid for all time slots. The inequality (29) proves the RHS of (27).
\end{proof}
We can use the upper and lower bounds on $Q(t)$, provided in (27), to find the appropriate values of $V$ and $C$ to make sure that the virtual queue $Q(t)$ remains stable during all the time slots. In the following corollary, we provide the possible values of parameters $V$ and $C$.
\begin{corollary}
The values of parameters $C$ and $V$ are given as follows
\bea \label{e6}
C=\Delta t P_{max}+E_{min}+\frac{V\gamma_{max}}{\Delta t},
\eea
\bea \label{e7}
V \in (0,V_{max}],~~~~~~~~~~~~
\eea
where
\bea
V_{max}=\frac{\Delta t(E_{max}-E_{min}-E_{c,max}-\Delta t P_{max})}{\gamma_{max}}.
\eea
\end{corollary}
\begin{proof}
See Proposition 2 in [10]. 
\end{proof}
By putting the value of $C$ from (\ref{e6}) in $Q(t)=E_b(t)-C$ and using the fact that $-\frac{V\gamma_{max}}{\Delta t}-\Delta t P_{max}\leq Q(t)\leq E_{c,max}$, we can easily show that $E_{min}\leq E_b(t)\leq E_{max}$ for all $t$. In the following Lemma we show that $P^{P6}(t)$ satisfies (6b) for all $t$.
\begin{lemma}
If we use $C=\Delta t P_{max}+E_{min}+\frac{V\gamma_{max}}{\Delta t}$ then $P^{P6}(t)$ satisfies the following constraint
\bea
\Delta t P^{P6}(t)\leq E_b(t)-E_{min},
\eea
for all $t$.
\end{lemma} 
\begin{proof}
In order to prove (33), first we show that $\Delta t P^{P7}(t)\leq E_b(t)-E_{min}$ is valid for all $t$. As we have shown above that $E_b(t)\geq E_{min}$ for all $t$, $\Delta t P^{P7}(t)\leq E_b(t)-E_{min}$ is valid for all $t$ with $P^{P7}(t)=0$. Hence, we only need to consider the case when $P^{P7}(t)>0$. From Lemma 3, we note that $P^{P7}(t)>0$ only when $Q(t)>-\frac{V\gamma_K(t)}{\Delta t}$. In this case, the inequality $\Delta t P^{P7}(t)\leq E_b(t)-E_{min}$ can be written as
\bea
\Delta t P^*(t)\leq Q(t)+C-E_{min}. 
\eea
Using the value of $C$ from (26), we can write (30) as
\bea
\Delta t P^*(t) \leq \frac{V \gamma_{max}}{\Delta t}+Q(t)+\Delta t P_{max}.
\eea
Since $\frac{V \gamma_{max}}{\Delta t}+Q(t)>0$ for $Q(t)>-\frac{V\gamma_K(t)}{\Delta t}>-\frac{V\gamma_{max}}{\Delta t}$ and $P^{P7}(t)\leq P_{max}$ we conclude that $\Delta t P^{P7}(t)\leq E_b(t)-E_{min}$ is satisfied for all $t$. According to Theorem 1, $P^{P6}(t)\leq P^{P7}(t)$ therefore we conclude that $\Delta t P^{P6}(t)\leq E_b(t)-E_{min}$ is also satisfied for all $t$. This completes the proof.
\end{proof}

\subsection{Non-Orthogonal Multiple Access with Individual User's Rate Constraints (NOMA-WR)}
With individual rate constraints, it is theoretically impossible to meet the rate requirements for all the time slots if we do not impose assumptions on the channel gains and the harvested energy. In this regard, we enlist following assumptions which are considered to be true in the following discussion.
\begin{itemize}
\item A1: First assumption is that $\gamma_i(t)\geq \gamma_{min}, \forall t, \forall i \in \{1,2,\cdots K\}$\footnote{Note that for simplicity of exposition we assume that all the users have same minimum bound on the channel gains however the forthcoming procedure can be adopted for the case where $\gamma_i(t)\geq \gamma_{min}^i$, where we may have $\gamma_{min}^i\neq \gamma_{min}^j, \forall i,j \in \{1,2,\cdots K\}$.}.
\item A2: Secondly, we assume that the arrived energy is higher than a certain threshold $E_{th}^{worst}$, where the value of $E_{th}^{worst}$ is found below.
\end{itemize}
The long-term time average sum rate maximization problem with the individual user's rate constraints can be formulated as
\begin{align}
\textbf{P8}
\max_{\{P(t), \boldsymbol{\rho}(t)\}}&~~~~\frac{1}{T}\lim_{T\rightarrow \infty}\sum_{t=0}^{T-1}\mathbb{E}\left(\sum_{k=1}^KR_k^{NOMA}(t)\right)\label{P1}\\
\text{s.t. }& \text{6a-6c, 6f}~~ \tag{36a} \nonumber \\
& R_k^{NOMA}(t)\geq R_{k,min}~\forall~ k \in \{1,\cdots K\},~~~\tag{36b}\nonumber\\
&\text{10b, 10c}, ~~\tag{36c} \nonumber
\end{align}
where $R_{k,min}$ is the minimum data rate requirement of the $k$-th user. Using (3), the constraint (36b) can be equivalently written as 
\bea
\rho_k(t) \geq M_k\left(\sum_{i=k+1}^K\rho_i(t)+\frac{1}{\gamma_k(t)P_{max}}\right),
\eea
where $M_k=\left(2^{R_{k,min}}-1\right)$. It can easily be shown that the constraint (36b) is satisfied only if $P_{max}\geq P_{th}(t)$, where $P_{th}(t)$ is given as 
\bea
P_{th}(t)=\sum_{k=1}^KP_{k,min}(t),
\eea
and $P_{k,min}(t)$ is given as
\bea
P_{k,min}(t)=M_k\left(\sum_{i=k+1}^KP_{i,min}(t)+\frac{1}{\gamma_k(t)}\right).
\eea
The value of $E_{th}^{worst}=\Delta t P_{th}^{worst}$ where $P_{th}^{worst}$ is obtained by putting $\gamma_i(t)=\gamma_{min}$ in (39) and using (38). It is straight forward to show that $P_{th}(t)\leq P_{th}^{worst}, \forall t$. 

After introducing the virtual queue $Q(t)=E_b(t)-C$ and following the similar procedure presented in the previous subsection, the per time slot optimization problem can be written as follows 
\begin{align}
\textbf{P9}
\min_{\{P(t), \boldsymbol{\rho}(t)\}} &Q(t)\left(E_h(t)-\Delta t P(t)\right)-V\sum_{k=1}^KR_k^{NOMA}(t),\label{P3}\\
&\text{s.t.~~~~} \text{6a, 10b, 10c, 36b}.\tag{40a} \nonumber
\end{align}
By introducing the following variables 
\bea
x_k(t)=\sum_{i=k+1}^K\rho_i(t), ~\forall ~ k \in \{1 \cdots K-1\},
\eea
\bea
\zeta(t)=\sum_{i=1}^K\rho_i(t)=\frac{P(t)}{P_{max}},
\eea
\bea
G_k(x_k(t))=\log(1+P_{max}\gamma_{k+1}(t)x_k(t))\nonumber \\-\log(1+P_{max}\gamma_{k}(t)x_k(t)),
\eea
we can write optimization problem \textbf{P9} as
\begin{align}
\textbf{P10}
\min_{\{\zeta(t), \boldsymbol{\rho}(t),\mathbf{x}(t)\}} -Q(t)\Delta t \zeta(t)P_{max}\nonumber \\-V\left(\log(1+\zeta(t)P_{max}\gamma_1(t))+\sum_{i=1}^{K-1}G_i(x_i(t)) \right)\label{P4}
\end{align}
\begin{align}
\text{s.t.} &&\zeta(t)\in\left[\frac{P_{th}}{P_{max}},1\right],\tag{44a}~~~~~~~~~~~~~~~~~~~ \nonumber 
\end{align}
\begin{align}
&&\zeta(t)=\sum_{i=1}^K\rho_i(t), \tag{44b}~~~~~~~~~~~~~~~~~~\nonumber
\end{align}
\begin{align}
&&~~~~~~~~~~~~x_k(t)=\sum_{i=k+1}^K\rho_i(t), ~\forall ~ k \in \{1 \cdots K-1\}, \tag{44c}\nonumber
\end{align}
\begin{align}
~~~~~~~~~~~&&\rho_k(t) \geq M_k\left(\sum_{i=k+1}^K\rho_i(t)+\frac{1}{\gamma_k(t)P_{max}}\right),\tag{44d} \nonumber 
\end{align}
where $\mathbf{x}(t)=\{x_1(t),\cdots, x_{K-1}(t)\}$. Problem \textbf{P10} can be decomposed into two optimization problems. First, we fix $\zeta(t)$ in the interval $[\frac{P_{th}(t)}{P_{max}},1]$ and minimize with respect to $\rho_k(t)$'s $\forall ~k \in \{1,\cdots K\}$ and then we minimize with respect to $\zeta(t)$.
\subsubsection{Minimization With Respect To $\rho_{i}(t)$'s}
For a fixed $\zeta(t)$, the optimization problem \textbf{P10} is equivalent to the following problem 
\begin{align}
\textbf{P11}
\max_{\{\boldsymbol{\rho}(t),\mathbf{x}(t)\}} \sum_{k=1}^{K-1} G_k(x_k(t)) \label{P5}
\end{align}
\begin{align}
\text{s.t.~~} \text{44b-44d},\tag{45a}~~~~~~~~~~ \nonumber
\end{align}

It can be easily verified that $G_k(x_k(t))$ is an increasing function of $x_i(t)$ $\forall ~i \in \{1,\cdots K-1\}$. Hence, maximizing $\sum_{i=1}^{K-1} G_i(x_i(t))$ is equivalent to maximizing individual $x_i(t)$'s while satisfying the constraints of \textbf{P10}. The solution of optimization problem \textbf{P11} is given in the following lemma.
\begin{lemma}
The optimal $\rho_i(t)$'s for a fixed value of $\zeta(t)\in [\frac{P_{th}(t)}{P_{max}}, 1]$ are given as

\bea
\rho_{k,\zeta(t)}^*(t)=N_k\left(\zeta(t)-\sum_{i=1}^{k-1}\rho_{i,\zeta(t)}^*(t)+\frac{1}{P_{max}\gamma_k(t)}\right),
\eea
if $k\neq K$ and 
\bea
\rho_{K,\zeta(t)}^*(t)=\zeta(t)-\sum_{i=1}^{K-1}\rho_{i,\zeta(t)}^*(t),
\eea
where $N_k=M_k/2^{R_{k,min}}$.
\end{lemma}
\begin{proof}
See theorem 2 of [22].
\end{proof}
According to lemma 1, the optimal power allocation has following properties.
\begin{itemize}
\item P1: The powers allocated to users $\{1,\cdots K-1\}$ are such that their data rate constraints are met with equality.
\item P2: The power allocated to $K$-th user is such that its data rate is $\geq R_{K,min}$.
\end{itemize}
We have the following corollary, which will be used later.
\begin{corollary}
The maximum sum rate is increasing in $\zeta(t)$ for $\zeta(t) \in \left[\frac{P_{th}(t)}{P_{max}},1\right]$. 
\end{corollary}
\begin{proof}
Consider $\zeta_1(t), \zeta_2(t)\in \left[\frac{P_{th}(t)}{P_{max}},1\right]$ and assume that $\zeta_1(t)>\zeta_2(t)$. Since $\zeta_1(t)$ and $\zeta_2(t)$ are both feasible, we can find optimal power allocation for both of them with the help of Lemma 1. Let us denote the power allocation obtained for $\zeta_1(t)$ by $\boldsymbol{\rho^1}(t)=\{\rho_{1,\zeta_1(t)}(t),\cdots \rho_{K,\zeta_1(t)}(t)\}$ and that obtained for $\zeta_2(t)$ by $\boldsymbol{\rho^2}(t)=\{\rho_{1,\zeta_2(t)}(t),\cdots \rho_{K,\zeta_2(t)}(t)\}$. According to property P1, the optimal sum rate for users $\{1,\cdots K-1\}$ will be same for both $\boldsymbol{\rho^1}(t)$ and $\boldsymbol{\rho^2}(t)$. On the other hand, since the derivative of $\rho_{K,\zeta(t)}(t)$ with respect to $\zeta(t)$ is $>0$ [22], the data rate achieved for the $K$-th user will be higher for $\boldsymbol{\rho^1}(t)$ as compared to that achieved for $\boldsymbol{\rho^2}(t)$. As a result the maximum sum rate obtained for $\zeta_1(t)$ will be higher than that achieved for $\zeta_2(t)$.  
\end{proof}
\subsubsection{Minimization with Respect To $\zeta(t)$}
After putting the optimal power allocation, obtained from Lemma 5, in the objective function of \textbf{P10} we get
\begin{align}
\textbf{P12}
\min_{\{\zeta(t)\}} -Q(t)\Delta t \zeta(t)P_{max}-V\bigg[\log(1+\zeta(t)P_{max}\gamma_1(t))\nonumber \\+\sum_{i=1}^{K-1}G_i(x_i^*(t)) \bigg]\label{P4}
\end{align}
\begin{align}
\text{s.t.~~~~} \text{44a}, \tag{48a} \nonumber
\end{align}
where $x_{k}^*(t)=\sum_{i=k+1}^K\rho_{i,\zeta(t)}^*(t)$. Note that $\rho_{i,\zeta (t)}^*(t)$ and $x_{i}^*(t)$ are affine functions of $\zeta(t)$. Additionally, the double derivative of $G_i(x_i(t))$ with respect to $x_i(t)$ is given as 
\bea
\frac{d^2G_i(x_i)}{dx_i^2}=\frac{P_{max}^2(\gamma_{i}-\gamma_{i+1})(2P_{max}\gamma_{i+1}\gamma_ix_i+\gamma_{i+1}+\gamma_{i})}{(P_{max}\gamma_{i+1}x_i+1)^2(P_{max}\gamma_ix_i+1)^2}. \nonumber
\eea
 As $\frac{d^2G_i(x_i)}{dx_i^2}<0$ and since convexity is preserved under affine transformation, we conclude that the objective function of \textbf{P12} is convex with respect to $\zeta(t)$. This implies problem \textbf{P12} is a convex optimization problem. Although the optimal $\zeta(t)$ does not admit a close form solution, we can use Bisection algorithm to find the solution of \textbf{P12}.
\subsubsection{Stabilizing the Virtual Queue}
Again, the non-close form solution of \textbf{P12} makes it difficult to find the appropriate values of $V$ and $C$. However, in the following we show that the same values of $V$ and $C$ that were obtained in subsection III-A can also be used for the problem in this subsection if the arrived energy is $\geq \Delta t P_{th}^{worst}$. In order to prove this statement, first we establish the bounds on the values of $Q(t)$ in the following Theorem.
\begin{theorem}
The value of $Q(t)$ is bounded as follows.
\bea \label{e2}
-\frac{V\gamma_{max}}{\Delta t}-\Delta tP_{max}\leq -\frac{V\gamma_{max}}{\Delta t}\leq Q(t) \leq E_{c,max}. 
\eea
\end{theorem}
\begin{proof}
First, we observe the following properties of the solution of problem \textbf{P7} and problem \textbf{P12}.
\begin{itemize}
\item P3: The optimal values of $\zeta(t)$, and in turn $P(t)=\zeta(t)P_{max}$, for problems \textbf{P7} and \textbf{P12} are non-decreasing with increasing values of $Q(t)$.
\item P4: When the optimal transmit power for problem \textbf{P7}, denoted by $P^{P7}(t)$, is zero, the optimal transmit power for the problem \textbf{P12}, denoted by $P^{P12}(t)$, is $P_{th}(t)$. (Note this property is different from the problem in the previous subsection.)
\item P5: There is a value of $Q(t)$, denoted by $Q^{crit}(t)$, for which the value of $P^{P7}(t)=P^{P12}(t)$.
\end{itemize}
Property P3 is straightforward to prove since the corresponding rate function is increasing function of $P(t)=\zeta(t)P_{max}$ and the slope of $-Q(t)P(t)$ decreases with increasing value of $Q(t)$. 

As in Theorem 1, it is easy to establish that for any fixed value of $P(t)=\zeta(t)P_{max}$
\bea
\log(1+P(t)\gamma_K(t)) \geq \log(1+\zeta(t)P_{max}\gamma_1(t))+\nonumber \\ \sum_{i=1}^{K-1}G_i(x_i^*(t)).
\eea
One easy way to prove (50) is to observe that, for any value of $P(t)=\zeta(t)P_{max}$, the LHS is the objective value achieved by the objective value if we relax the data rate constraints by replacing $R_{k,min}=0, \forall k\in \{1,2,\cdots K\}$. This means that if the derivative of the objective function of \textbf{P7} with respect to $P(t)$ has become positive, then the derivative of the objective function of \textbf{P12} is definitely positive. Such behavior implies that the optimal transmit power for \textbf{P12} should be equal to the lowest possible feasible value, which in this problem is $P_{th}(t)$. Hence, property P4 is also proved. 

For a certain range of $Q(t)$, the value of the optimal transmit power for problem \textbf{P7} is smaller than the optimal transmit power for problem \textbf{P12}. However, as $Q(t)$ increase the value of the optimal transmit power for problem \textbf{P7} will increase and for specific value of $Q(t)=Q^{th}(t)$ it will become equal to the optimal transmit power for problem \textbf{P12}. Now, using the result from Lemma 3 we can have the following equality
\bea
-\frac{V}{\Delta t Q^{th}(t)}-\frac{1}{\gamma_K(t)}=P^{P12}(t),
\eea
which is obtained by equating the value of optimal transmit power for problem \textbf{P7} from Lemma 3 to the optimal transmit power of problem \textbf{P12}. As the minimum value of $P^{P12}(t)$ cannot be smaller than $P_{th}^{best}$, which is the minimum required transmit power for satisfying the data rate requirements when all the channel gains are equal to $\gamma_{max}$, and $\gamma_k(t)\leq \gamma_{max}, \forall k \in\{1,2,\cdots, K\}$, we have the following inequality
\bea
Q^{th}(t)>Q^{crit}=-\frac{V}{\Delta t}\left(\frac{1}{P_{th}^{best}+\frac{1}{\gamma_{max}}}\right).
\eea
Now, assume that in a certain time slot $-\frac{V\gamma_{max}}{\Delta t} \leq Q(t)\leq Q^{crit}$ then the optimal transmit power for the proposed scheme is $P_{th}(t)$. This is because the objective function is increasing in $P(t)$ for all the feasible values of $P(t)$. Hence, the minimum of the objective is achieved when $P(t)$ is set equal to its lowest feasible value i.e. $P_{th}(t)$. This implies that $Q(t)$ is always increasing in next time slot because $P_{th}(t)\leq P_{th}^{worst}$.

On the other hand, if $Q(t)>Q^{crit}(t)$, then we have $P(t)>P_{th}(t)$. However, if we choose 
\begin{align}
Q^{crit}-\left(-\frac{V\gamma_{max}}{\Delta t}\right)=&\frac{V\gamma_{max}}{\Delta t}\left(\frac{\gamma_{max}P_{th}^{best}}{1+\gamma_{max}P_{th}^{best}}\right)\nonumber \\ \geq & \Delta t P_{max}, \nonumber
\end{align}
then $Q(t)$ will always be greater than $-\frac{V\gamma_{max}}{\Delta t}$ in the next time slot\footnote{Note that if we chose $V=\frac{\Delta t (E_{max}-E_{min}-E_{c,max}-\Delta t P_{max})}{\gamma_{max}}$. Then, the condition $\frac{V\gamma_{max}}{\Delta t}\left(\frac{\gamma_{max}P_{th}^{best}}{1+\gamma_{max}P_{th}^{best}}\right) \geq \Delta t P_{max}$ translate into $E_{max}-E_{min}-E_{c,max}-\Delta t P_{max}\geq \Delta t P_{max}\left(\frac{1+\gamma_{max}P_{th}^{best}}{\gamma_{max}P_{th}^{best}}\right)$.}. Hence, we conclude that
\bea
Q(t)\geq -\frac{V\gamma_{max}}{\Delta t} \geq -\frac{V\gamma_{max}}{\Delta t}-\Delta t P_{max},
\eea
is a valid bound.

In a similar way, by checking the possibilities $Q(t)>0$, $Q(t)\leq 0$ and using the fact that $E_{c,max}\leq \Delta t P_{max}$ we can verify that 
\bea \label{e1}
Q(t+1)\leq Q(t)+E_{c,max}\leq E_{c,max},
\eea  
is valid for all time slots. The inequality (\ref{e1}) proves the RHS of (\ref{e2}).
\end{proof}
Since the $Q(t)$ has similar bounds as in the previous subsection, we can easily show the following values of $V$ and $C$ 
\bea
C=\Delta t P_{max}+E_{min}+\frac{V\gamma_{max}}{\Delta t},
\eea
\bea
V=\frac{\Delta t (E_{max}-E_{min}-E_{c,max}-\Delta t P_{max})}{\gamma_{max}},
\eea
make sure that battery level remains in the prescribed limits during all the time slots i.e. $E_{min}\leq E_b(t)\leq E_{max}$. 

Next, we have to show that the battery has enough energy to be able to transmit $P^{P12}(t)$. This is proved in the following lemma.
\begin{lemma}
It is always feasible for battery to transmit $P^{P12}(t)$ in the proposed scheme.
\end{lemma}
\begin{proof}
We know that $P^{P12}(t)$ is feasible only if the battery has enough energy to support transmission of $P^{P12}(t)$. From Theorem 2, we also know that $Q(t)\geq -\frac{V\gamma_{max}}{\Delta t}$. Furthermore, by putting the value of $C$ in $E_{b}(t)=Q(t)+C$ and noting that $P^{P12}(t) \leq P_{max}$ we prove that it is always feasible for battery to transmit $P^{P12}(t)$ in the proposed scheme.
\end{proof}
\section{Online Rate Maximizing Power Allocation for Orthogonal Multiple Access}

First, we solve the long-term sum rate maximization problem when the individual users do not have any minimum rate requirements. Next, we solve the long-term sum rate maximization problem when the individual users have minimum rate requirements which should be met for all the time slots.  
\subsection{Orthogonal Multiple Access without Individual User Rate Constraints (OMA-WoR)}

After introducing the virtual queue $Q(t) = E_b(t) - C$ and following the procedure described in Section III-A, the per time slot optimization problem can be written as follows
\begin{align}
\textbf{P13}
\min_{\{\boldsymbol{\rho}(t),\boldsymbol{\alpha}(t),P(t)\}}&~ Q(t)\left(E_h(t)-\Delta t P(t)\right)~~~~\nonumber \\&-V\sum_{k=1}^KR_k^{OMA}(t),\\
\text{s.t.~~~~ }&\text{6a} \tag{57a} \nonumber \\
&\text{10b-10e}. \tag{57b}\nonumber
\end{align}
By introducing new variables 
\bea
P_k(t)=\rho_k(t)P(t), \forall k \in \{1,2,\cdots, K\},
\eea
the above problem can be transformed into following equivalent problem
\begin{align}
\textbf{P14}
\min_{\{\mathbf{P}(t),\boldsymbol{\alpha}(t),P(t)\}}& Q(t)\left(E_h(t)-\Delta t P(t)\right)\nonumber \\&-V\sum_{k=1}^KG_k^{OMA}(P_k(t),\alpha_k(t)),
\end{align}
\begin{align}
\text{s.t.~~~~~~ }&0\leq \sum_{k=1}^KP_k(t)=P(t) \leq P_{max}, \tag{59a} \nonumber\\
&0\leq P_k(t) \leq P(t),\tag{59b}\nonumber \\
& \text{10d,10e}.\tag{59c} \nonumber
\end{align}
where 
\bea
G_k^{OMA}(P_k(t),\alpha_k(t))=\alpha_k(t)\log\left(1+\frac{P_k(t)\gamma_k(t)}{\alpha_k(t)}\right),
\eea
and $\mathbf{P}(t)=\{P_1(t),\cdots, P_{K}(t)\}$. It can be easily shown that the objective function of the \textbf{P14} is jointly convex in $\{\mathbf{P}(t),\boldsymbol{\alpha}(t)\}$ and therefore the above problem is a convex optimization problem. We have the following lemma for problem \textbf{P14}. 
\begin{lemma}
The problem \textbf{P14} is equivalent to the problem \textbf{P7} and the optimal solutions for $\alpha_k(t),\rho_k(t),P_k(t) \forall k \in\{1,2,\cdots K-1\}$ are given as follows 
\bea
\alpha_k^*(t)=\rho_k^*(t)=P_k^*(t)=0,~~ \forall~~ k \in \{1,2,\cdots K-1\}.
\eea
\end{lemma}
\begin{proof}
For any fixed value of $P(t)$, the problem \textbf{P14} can be written as
\begin{align}
\textbf{P15}
\max_{\{\boldsymbol{\alpha}(t),\mathbf{P}(t)\}} &\sum_{k=1}^KG_k^{OMA}(P_k(t),\alpha_k(t))
\end{align}
\begin{align}
\text{s.t. }&\sum_{k=1}^KP_k(t)\leq P(t), \tag{62a} \nonumber \\
&\sum_{k=1}^K\alpha_k(t)\leq 1.\tag{62b} \nonumber
\end{align}
This is because $G_k^{OMA}(P_k(t),\alpha_k(t))$'s are increasing functions of $P_k(t),\alpha_k(t)$. Consider the case of $K=2$ and $ \alpha_1(t)=1-\alpha_2(t), \alpha_2(t) \leq 1$ then \textbf{P15} can be equivalently written as 
\begin{align}
\textbf{P16}
\max_{\{\alpha_2(t),P_2(t)\}} (1-\alpha_2(t))&\log\left(1+\frac{\gamma_{1}(t)(P(t)-P_2(t))}{1-\alpha_2(t)}\right)\nonumber \\& +\log\left(1+\frac{\gamma_2(t)P_2(t)}{\alpha_2(t)}\right),
\end{align}
\begin{align}
\text{s.t.   }&P_2(t) \leq P(t), \tag{63a}\nonumber\\
&\alpha_2(t) \leq 1.\tag{63b}\nonumber
\end{align}
The problem \textbf{P16} is convex and hence Karush-Kuhn Tucker (KKT) conditions are sufficient for finding the solution of problem \textbf{P16}. The KKT conditions in the case of inactive constraints can be written as
\bea \label{e3}
\frac{\gamma_2(t)\alpha_2^*(t)}{\alpha_2^*(t)+\gamma_2(t)P_2^*(t)}=\frac{\gamma_1(t)(1-\alpha_2^*(t))}{1-\alpha_2^*(t)+\gamma_1(t)(P(t)-P_2^*(t))},
\eea
\bea \label{e4}
&\log\left(1+\frac{\gamma_2(t)P_2^*(t)}{\alpha_2^*(t)}\right)+\frac{\gamma_1(t)(P(t)-P_2^*(t))}{1-\alpha_2^*(t)+\gamma_1(t)(P(t)-P_2^*(t))}\nonumber \\&=\log\left(1+\frac{\gamma_1(t)(P(t)-P_2^*(t))}{1-\alpha_2^*(t)}\right)+\frac{\gamma_2(t)P_2^*(t)}{\alpha_2^*(t)+\gamma_2(t)P_2^*(t)}.
\eea
Since the LHS of (\ref{e3}) is an increasing function of $\frac{\gamma_2(t)P_2^*(t)}{\alpha_2^*(t)}$ and the RHS of (\ref{e3}) is an increasing function of $\frac{\gamma_1(t)(P(t)-P_2^*(t))}{1-\alpha_2^*(t)}$, we must have $\frac{\gamma_2(t)P_2^*(t)}{\alpha_2^*(t)}=\frac{\gamma_1(t)(P(t)-P_2^*(t))}{1-\alpha_2^*(t)}$. But combining this fact with (\ref{e4}) we must have $\gamma_2(t)=\gamma_1(t)$ which contradicts with the fact that $\gamma_2(t)>\gamma_1(t)$. Therefore, we conclude that at least one of the constraint must be active in problem \textbf{P16}. Now assuming that if $P_2^*(t)=P(t)$, then we must have $\alpha_2(t)=1$. Similarly, if we assume that $\alpha_2(t)=1$, then we must have $P_2^*(t)=P(t)$ for optimality. Same reasoning can also be used to establish the result for more than two users.  
\end{proof}
Hence, the problem \textbf{P14} translates to a single user problem and its solution is presented in Lemma 3. Therefore, we impose a constraint on the bandwidth and power allocation as follows
\bea
\alpha_1(t)\geq \alpha_2(t)\geq \cdots \geq \alpha_K(t), \\
\rho_1(t)\geq \rho_2(t) \geq \cdots \geq \rho_K(t).
\eea
With these constraints added to the problem \textbf{P14}, we can write the multiple user optimization problem as follows.
\begin{align}
\textbf{P17}
\min_{\{\boldsymbol{\alpha}(t),\mathbf{P}(t),P(t)\}}& Q(t)\left(E_h(t)-\Delta t P(t)\right)\nonumber \\&-V\sum_{k=1}^KG_k^{OMA}(P_k(t),\alpha_k(t)),
\end{align}
\begin{align}
\text{s.t. }&\text{59a, 59b} \tag{68a} \nonumber \\
& \text{10d, 10e},\tag{68b} \nonumber \\
&\text{66, 67}. \tag{68c}\nonumber 
\end{align}
Since the inclusion of constraints (68c) does not affect the convexity of the problem \textbf{P14}, we can easily obtain the solution of the problem with the newly added constraints. For a fixed value of $P(t)$, $G_k^{OMA}(P_k(t),\alpha_k(t))$'s are increasing functions of $P_k(t),\alpha_k(t)$. Therefore, the optimal solution for $\alpha_k(t)$'s are $\frac{1}{K}$ and $P_k(t)$'s are $\frac{P(t)}{K}$ for a fixed value of $P(t)\in \{0,P_{max}\}$. Additionally, we can follow the procedure outlined in Theorem 1 to show that the optimal transmit power of the problem \textbf{P17} is always smaller than the optimal transmit power for \textbf{P7}. Hence, it follows that the $Q(t)$ is bounded as described in (\ref{e5}). Subsequently, it follows that the same values of $V, C$ given in (\ref{e6}), (\ref{e7}) can be used in this problem.
\subsection{Orthogonal Multiple Access with Individual User's Rate Constraints (OMA-WR)} 
The long-term time average sum rate maximization problem with the individual user’s rate constraints can be formulated as
\begin{align}
\textbf{P18}
\max_{\{P(t), \boldsymbol{\rho}(t),\boldsymbol{\alpha}(t)\}}&\frac{1}{T}\lim_{T\rightarrow \infty}\sum_{t=0}^{T-1}\mathbb{E}\left(\sum_{k=1}^KR_k^{OMA}(t)\right)\label{P1}\\
\text{s.t. }&\text{6a-6c, 6f,} \tag{69a}, \nonumber \\
& \text{10b-10e}, \tag{69b} \nonumber \\
& R_k^{OMA}(t)\geq R_{k,min}.~~~\tag{69c}\nonumber
\end{align}
Again, we assume that assumptions A1 and A2 are applicable. Following the procedure of previous section, the per time slot based optimization problem can be written as follows 
\begin{align}
\textbf{P19}
\min_{\{P(t),\mathbf{P}(t),\boldsymbol{\alpha}(t)\}}& Q(t)\left(E_h(t)-\Delta t P(t)\right)\nonumber \\&-V\sum_{k=1}^KG_k^{OMA}(P_k(t),\alpha_k(t)),
\end{align}
\begin{align}
\text{s.t. }&\text{59a, 59b}, \tag{70a} \nonumber \\
& \text{10d, 10e,} \tag{70b} \nonumber \\
& G_k^{OMA}(P_k(t),\alpha_k(t))\geq R_{k,min}. \tag{70c} \nonumber
\end{align}
The minimum transmit power required to satisfy the individual rate constraints can be obtained by solving the following problem
\begin{align}
\textbf{P20}
 \min_{\{\mathbf{P}(t),\boldsymbol{\alpha}(t)\}}& \sum_{k=1}^K P_k(t),
\end{align}
\begin{align}
\text{s.t. }& G_k^{OMA}(P_k(t),\alpha_k(t))\geq R_{k,min}, \tag{71a} \nonumber \\
&\sum_{k=1}^KP_k(t)\leq P_{max}, \tag{71b} \nonumber \\
&\sum_{k=1}^K \alpha_k(t)\leq 1. \tag{71c} \nonumber
\end{align}
It is easy to show that in the optimal solution the rate constraint and bandwidth constraint should be met with equality. Let $\beta$, $\theta$ and $\lambda_k$ denote the Lagrange multipliers for bandwidth, power allocation and k-th user rate constraints, respectively. Then, using the KKT conditions we can show that the dual variables $(\lambda_k,\beta,\theta)$ and primal variables $(\alpha_k(t),P_k(t))$ are related as follows
\bea
R_{k,min}=\alpha_k(t)\log(\lambda_k\gamma_k(t)),
\eea
\bea
\theta=\lambda_k\left(\log(\lambda_k\gamma_k(t))-1\right)+\frac{1}{\gamma_k(t)},
\eea
\bea
\frac{P_k(t)}{\alpha_k(t)}=\lambda_k-\frac{1}{\gamma_k(t)},
\eea
\bea
\sum_{k=1}^K\alpha_k(t)=1.
\eea
We denote the optimal objective value achieved in problem \textbf{P20} by $P_{th}^{OMA}(t)$. Then, the problem \textbf{P19} can be equivalently written as follows
\begin{align}
\textbf{P21}
\min_{\{P(t),\mathbf{P}(t),\boldsymbol{\alpha}(t)\}}& Q(t)\left(E_h(t)-\Delta t P(t)\right)\nonumber \\&-V\sum_{k=1}^KG_k^{OMA}(P_k(t),\alpha_k(t)),
\end{align}
\begin{align}
\text{s.t. }&P_{th}^{orth}(t) \leq \sum_{k=1}^KP_k(t)=P(t) \leq P_{max}, \tag{76a} \nonumber\\
&\text{10d, 10e}, \tag{76b} \nonumber \\ 
&\text{59b},\tag{76c} \nonumber \\
& \text{70c}. \tag{76d} \nonumber 
\end{align}
For a fixed value of $P(t)\in \{P_{th}^{OMA}(t),P_{max}\}$, the problem \textbf{P21} can be equivalently written as 
\begin{align}
\textbf{P22}
\max_{\{\mathbf{P}(t),\boldsymbol{\alpha}(t)\}}\sum_{k=1}^KG_k^{OMA}(P_k(t),\alpha_k(t)),
\end{align}
\begin{align}
\text{s.t. }&\text{71a-71c}. \tag{77a} \nonumber
\end{align}
We have the following Lemma for problem \textbf{P22}.
\begin{lemma}
The optimal values of $\alpha_k(t),P_k(t)~~ \forall k\in \{1,2,\cdots K-1\}$ can be obtained from the following relation
\bea
\alpha_k(t)=\frac{R_{k,min}}{\log(1+g_k(t))}, ~~ P_k(t)=\frac{g_k(t)\alpha_k(t)}{\gamma_k(t)},
\eea
and for $\alpha_K(t),P_K(t)$ we have 
\bea \label{e15}
\alpha_K(t)=1-\sum_{k=1}^{K-1}\alpha_k(t), ~~P_K(t)=P(t)-\sum_{k=1}^{K-1}P_k(t),
\eea
where $g_k(t)=\frac{P_k(t)\gamma_k(t)}{\alpha_k(t)}$. The value of $g_k(t), \forall k \in\{1,2,\cdots K-1\}$ is given as follows
\bea \label{e14}
g_k(t)=\frac{\exp\left(W\left(\gamma_k(t)f(g_K(t))-1\right)\right)+1}{\gamma_k(t)},
\eea
where $W(x)$ is the Lambert-W function and $f(g_K(t))$ is given below 
\bea \label{e8}
f(g_K(t))=\frac{\log(1+g_K(t)\gamma_K(t))-\frac{1}{1+\frac{1}{g_K(t)\gamma_K(t)}}}{\theta}.
\eea
In (\ref{e8}), $\theta$ is the Lagrange multiplier associated with the power constraint in problem \textbf{P22} and $g_K(t)$ for a fixed value of $\theta$ is given as
\bea
g_K(t)=\frac{1}{\theta}-\frac{1}{\gamma_K(t)}.
\eea
\end{lemma}
\begin{proof}
The Lagrangian of the problem \textbf{P22} is given as
\bea
-\sum_{k=1}^K(\lambda_k+1)\alpha_k(t)\log\left(1+\frac{P_k(t)\gamma_k(t)}{\alpha_k(t)}\right)+\sum_{k=1}^K\lambda_kR_{k,min} \nonumber
\eea
\bea
+\beta \left(\sum_{k=1}^K\alpha_k(t)-1\right)+\theta \left(\sum_{k=1}^KP_k(t)-P(t)\right),
\eea
where $\beta,\theta$ and $\lambda_k$ are the Lagrange multipliers for bandwidth, power allocation and $k$-th user rate constraints, respectively. The corresponding KKT conditions for problem \textbf{P22} can be written as
\bea \label{e9}
\theta=(1+\lambda_k)\left(\frac{\gamma_k(t)\alpha_k(t)}{\alpha_k(t)+P_k(t)\gamma_k(t)}\right),
\eea
\bea \label{e10}
\beta=(1+\lambda_k)\bigg(\log\left(1+\frac{P_k(t)\gamma_k(t)}{\alpha_k(t)}\right) \nonumber \\-\frac{P_k(t)\gamma_k(t)}{\alpha_k(t)+P_k(t)\gamma_k(t)}\bigg),
\eea
\bea
\lambda_k\left(R_{k,min}-\alpha_k(t)\log\left(1+\frac{P_k(t)\gamma_k(t)}{\alpha_k(t)}\right)\right)=0,
\eea
\bea
\beta(\sum_{k=1}^K\alpha_k(t)-1)=0,
\eea
\bea
\theta(\sum_{k=1}^KP_k(t)-P(t))=0,
\eea
\bea
\beta \geq 0,~~\theta \geq 0,~~ \lambda_k\geq 0,~~ \forall k\in \{1,2,\cdots, K\}.
\eea
Since $P_k(t)>0,\alpha_k(t)>0$, we conclude from (\ref{e9}), (\ref{e10}) that $\beta>0$ and $\theta>0$. Thus, according to complementary slackness conditions power and bandwidth allocation constraints must be active. 

Next, we show that the data rate constraints for $\{1,2,\cdots, K-1\}$ must be active while that for $K$-th user can be inactive if $P(t)>P_{th}^{orth}(t)$. We prove this claim for two users, i.e. $K=2$, however same reasoning can be used for any number of users. We have following properties of $\lambda_1,\lambda_2.$ 
\begin{itemize}
\item P6: First, we note that $\lambda_1$ and $\lambda_2$ cannot be positive simultaneously. This is because if $\lambda_1,\lambda_2>0$, then according to the complementary slackness condition the rate constraints should be met with equality at optimality. However, if $P(t)>P_{th}(t)$, for same bandwidth allocation we can allocate excess power to the stronger user to increase its rate to increase the objective value. This contradicts the assumption that for optimality both the rate constraints should be met with equality. Hence, $\lambda_1$, $\lambda_2$ cannot be simultaneously greater than zero.
\item P7: $\lambda_1,\lambda_2$ cannot be zero simultaneously. This can be proved through contradiction. Assume $\lambda_1=0$ and $\lambda_2=0$. Then, from (\ref{e9}), we obtain
\bea \label{e11}
\frac{\theta}{\gamma_k(t)}=\frac{1}{1+\frac{P_k(t)\gamma_k(t)}{\alpha_k(t)}},~~ \forall k\in \{1,2\}.
\eea
Using (\ref{e9}), we can write (\ref{e10}) for $k\in \{1,2\}$ as follows
\bea \label{e12}
\beta=\log\left(\frac{\gamma_1(t)}{\theta}\right)-1+\frac{\theta}{\gamma_1(t)}\nonumber \\
=\log\left(\frac{\gamma_2(t)}{\theta}\right)-1+\frac{\theta}{\gamma_2(t)}.
\eea
From (\ref{e11}), we know that $\frac{\gamma_k(t)}{\theta}>1$, and we know that $\log(x)+\frac{1}{x}$ is an increasing function of $x$ for $x>1$. Thus, (\ref{e12}) cannot be satisfied unless $\gamma_1(t)=\gamma_2(t)$. However, this contradicts the assumption that $\gamma_2(t)>\gamma_1(t)$. Hence, $\lambda_1$ and $\lambda_2$ cannot be simultaneously zero. 
\item P8: $\lambda_2>0$, $\lambda_1=0$ is not possible. Since if we assume $\lambda_2>0,\lambda_1=0$, then we have
\bea
\frac{\theta}{\gamma_1(t)}=\frac{1}{1+\frac{P_1(t)\gamma_1(t)}{\alpha_1(t)}},
\eea
\bea
\frac{\theta}{(1+\lambda_2)\gamma_2(t)}=\frac{1}{1+\frac{P_2(t)\gamma_2(t)}{\alpha_2(t)}}.
\eea
Using (\ref{e10}), we can write $\beta$ as
\bea
\beta&=(1+\lambda_2)\left(\log\left(\frac{\gamma_2(t)(1+\lambda_2)}{\theta}\right)-1+\frac{\theta}{(1+\lambda_2)\gamma_2(t)}\right),\nonumber \\
&=\log\left(\frac{\gamma_1(t)}{\theta}\right)-1+\frac{\theta}{\gamma_1(t)}.
\eea
However, this leads to a contradiction since this is only possible if $\lambda_2=0$ and $\gamma_2(t)=\gamma_1(t)$. 
\end{itemize}
Similar reasoning can be used to show that only the $K$-th user rate constraint can be inactive for any integer value of $K>2$. Thus, we have $\lambda_K=0$ and (\ref{e9}), (\ref{e10}) can be arranged in following way
\bea
\beta=\log\left(1+\gamma_K(t)g_K(t)\right)-\frac{1}{1+\frac{1}{\gamma_K(t)g_K(t)}},
\eea 

\bea \label{e13}
\left(\frac{1}{\gamma_k(t)}+g_k(t)\right)\left(\log\left(1+g_k(t)\gamma_k(t)\right)\right)-g_k(t)\nonumber \\=\frac{\log(1+g_K(t)\gamma_K(t))-\frac{1}{1+\frac{1}{g_K(t)\gamma_K(t)}}}{\theta},
\eea
where $g_K(t)$ for a fixed value of $\theta$ can be obtained from the relation 
\bea
\theta=\frac{\gamma_{K}(t)}{1+g_K(t)\gamma_K(t)}.
\eea
From (\ref{e13}) we can easily obtain (\ref{e14}). Subsequently, we can use the value of $g_k(t), \forall k\in\{1,2,\cdots, K-1\}$ to obtain $\alpha_k(t)$ from $R_{k,min}=\alpha_k(t)\log\left(1+g_k(t)\right)$, and $P_k(t)$ can be obtained from $g_k(t)=\frac{P_k(t)\gamma_k(t)}{\alpha_k(t)}$. Then, the remaining bandwidth/power can be allocated to the $K$-th user which is mathematically represented by (\ref{e15}). 
\end{proof}

For each fix value of $P(t)$, the optimal value of $\theta$ can be obtained by employing a linear search. Although the optimal solution for \textbf{P21} cannot be obtained in closed form, by using similar reasoning presented in Section III, we can show that the values of $V$and $C$ presented in (\ref{e6}), (\ref{e7}) can also be used for this scenario to make sure that the battery level constraints are met for all the time slots.   

The proposed schemes for NOMA-WoR, NOMA-WR and OMA-WoR only require one Bisection search to find the optimal values of transmit power $P(t)$. For OMA-WR, linear search on two variables is required. On the contrary, dynamic programming require transition probabilities knowledge of the system states. Furthermore, the computational complexity of the DP is very high. For example, if the arriving energy, channel gains and total transmit power can be discretized into $J,H,I$ number of states, respectively. Then, the computational complexity of the DP for a time horizon $T$ is $\mathcal{O}(J^2\times H^2\times I^2\times T^2)$ [23]. In addition, the DP scheme requires the knowledge of state transition probabilities which is impossible to attain in a practical system.
\section{Performance Gap Analysis}
In this section, we present a theoretical analysis of performance comparison between the optimal solution and the proposed solution for the case of NOMA without individual user's rate constraint. However, the analysis can be easily extended to the other cases and thus will be omitted for brevity. It is well known that there exists a stationary, randomized power control policy $\{P_r(t)\}$ for problem \textbf{P3}, where $P_r(t)$ only depends on the current system state. For this policy, if we denote the per time slot sum rate by $\mathbb{G}_K^r(t)$, we have the following properties
\bea
\mathbb{E}[\mathbb{G}_K^r(t)]\triangleq \bar{\mathbb{G}}_K^r=\bar{\mathbb{G}}^o, \\
\mathbb{E}[E_h^r(t)]=\mathbb{E}[\Delta t P_r(t)],
\eea
where $\bar{\mathbb{G}}_K^r, \bar{\mathbb{G}}^o$ are the objective values achieved for problem \textbf{P3} under the policy $P_r(t)$ and the optimal policy for solving \textbf{P3}. As the proposed scheme minimizes the RHS of (16) on per time slot basis, we can write
\begin{align}
&\Delta (Q(t))-V \mathbb{E}[\mathbb{G}_K^p(t)|Q(t)] \nonumber \\ &\leq \phi+X(t)\mathbb{E}[E_h^r(t)-\Delta t P_r(t)|X(t)]-V\mathbb{E}[G_K^r(t)|Q(t)],
\end{align} 
where $\mathbb{G}_K^p(t)$ denotes the per time slot sum rate achieved by the proposed power allocation scheme. Since the randomized policy only depend on the current system state, we can write
\begin{align}
& \Delta (Q(t))-V \mathbb{E}[G_K^p(t)|Q(t)]\nonumber \\ &\leq \phi+X(t)\mathbb{E}[E_h^r(t)-\Delta t P_r(t)]-V\mathbb{E}[G_K^r(t)],
\end{align}
which can be further simplified by using (99) as 
\bea
\Delta (Q(t))-V \mathbb{E}[G_K^p(t)|Q(t)]\leq \phi-V\mathbb{E}[G_K^r(t)].
\eea
Using (98) we can write (102) as 
\bea
\Delta (Q(t))-V \mathbb{E}[G_K^p(t)|Q(t)] \leq \phi -V\bar{\mathbb{G}}_K^o.
\eea
Recall that \textbf{P2} is a relaxed problem of \textbf{P1} and therefore if we denote the optimal objective achieved by \textbf{P1} as $\bar{\mathbb{G}}_K^{opt}$ then we have
\bea
\Delta (Q(t))-V \mathbb{E}[G_K^p(t)|Q(t)] \leq \phi -V\bar{\mathbb{G}}_K^{opt},
\eea
where we have used that fact that optimal value achieved by a relaxed problem is always greater than the optimal value of the original problem hence $\bar{\mathbb{G}}_K^o\geq \bar{\mathbb{G}}_K^{opt}$. Taking the expectation with respect to $Q(t)$ on both sides and summing from $t=0$ to $t=T-1$, we have
\begin{align}
&V\sum_{t=0}^{T-1}\mathbb{E}[G_K^p(t)]\nonumber \\ &\geq TV \bar{\mathbb{G}}_K^{opt}-T\phi +\mathbb{E}[L(Q(T))]-\mathbb{E}[L(X(0))]
\end{align}
Diving both sides by $T$ and taking the limit $T\to \infty$, we get
\bea
\bar{\mathbb{G}}_K^{opt}-\lim_{T\to \infty}\frac{1}{T}\sum_{t=0}^{T-1}\mathbb{E}[G_K^p(t)]\leq \frac{\phi}{V}. 
\eea 
This implies that the performance gap between the optimal scheme and the proposed scheme is bounded, and this gap reduces with the increasing value of $V$. Hence, we conclude that by increasing the value of $V$ parameter the performance of the proposed schemes improves.
\color{black}
\section{Simulation Results}
In this section, we provide simulation results to illustrate the performance of the proposed scheme. The simulation parameters are given in Table 1. The superscripts WR and WoR indicates the parameters used for solving the problem with individual user rate requirements and without individual user rate requirements. It is assumed that $E_a(t)$ follows a compound Poisson process with uniform distribution. The arrival rate is denoted by $\lambda$ while the parameter for uniform distribution is denoted by $\alpha$. The channel gains are assumed to be distributed according to either exponential distribution or uniform distribution with parameter $1$. 

For NOMA-WoR scenario, we compare our results with two baseline schemes. In the first baseline scheme, the total transmit power is randomly distributed among the users while satisfying the power order constraint $\rho_1(t)\geq \rho_2(t)\cdots \geq \rho_K(t)$. We name this scheme as the random power allocation (RPA) scheme. In the second scheme, the total transmit power is equally distributed among all the users. This scheme is termed as the optimal power allocation (OPA) scheme. In both of the baseline schemes, the total transmit power is equal to $\min\{(E_b(t)-E_{min})/\Delta t,P_{max}\}$.  
\begin{table}[h]
\begin{center}
\caption{Simulation Parameters}
\begin{tabular}{|c|c|c|c|}
\hline 
Parameter & Value & Parameter & Value \\
\hline
\hline
$E_{max}$ & $10$J & $E_{min}$ & $0$J \\
\hline
$E_b(0)$ & $[0,E_{max}/2]$ & $\alpha$ & $.2$ \\
\hline
$\gamma_{max}$ & $20$dB & $\lambda$ & $[.5\sim 2.5]$ \\
\hline 
$\gamma_{min}$ & $10$dB & $R_{k,min}, \forall k \in\{1,2,\cdots K\}$ & $1$bps \\
\hline
$K$ & $4$ & $\Delta t$ & $1$s\\
\hline
$E_{c,max}^{WoR}$ & $.5$J & $P_{max}^{WoR}$ & $1$W\\
\hline
$E_{c,max}^{WR}$ & $1.6$J & $P_{max}^{WR}$ & $2$W\\
\hline 
\end{tabular}
\end{center}
\end{table}

Fig. 3 and Fig. 4 show the time averaged throughput for the three schemes for exponentially distributed channel gains and uniformly distributed channel gains, respectively. The time averaged sum rate for all the schemes converges to the same value for both initial statuses of the battery. It is clear that the proposed scheme outperforms the baseline schemes in the long run for both scenarios of channel gains distribution. Also, the power allocation among users can significantly affect
\begin{figure}[h]
\begin{center}
\includegraphics[width=3.5in]{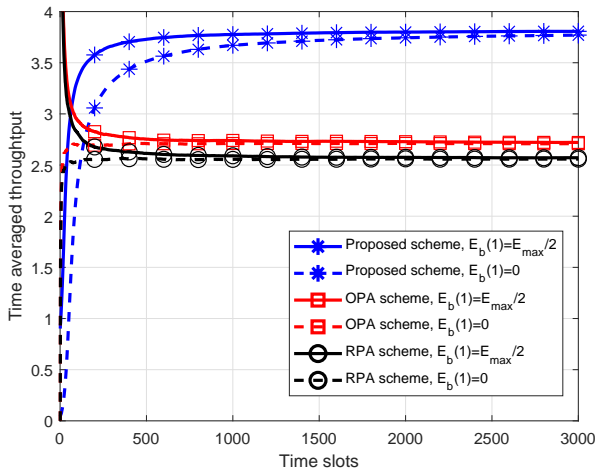}
\end{center}
\caption{Time averaged throughput with respect to time slots with when channel gains are exponentially distributed for NOMA-WoR scenario.}
\end{figure}
\begin{figure}[h]
\begin{center}
\includegraphics[width=3.5in]{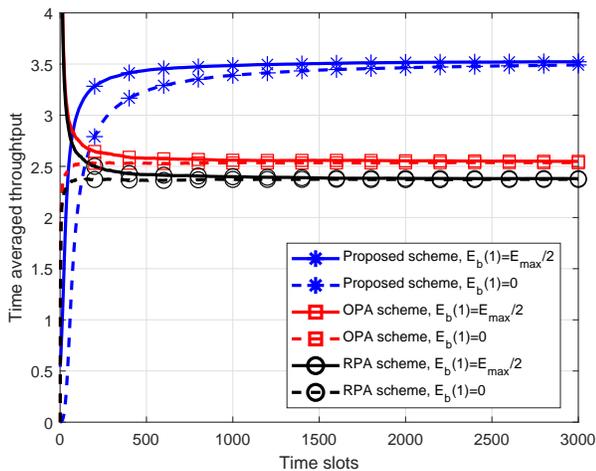}
\end{center}
\caption{Time averaged throughput with respect to time slots with when channel gains are uniformly distributed for NOMA-WoR scenario.}
\end{figure}
the performance and is reflected by the superior performance of OPA with respect to RPA. The improvement in performance of the proposed scheme as compared to the baseline schemes is explained as follows. First, the total transmit power is optimally allocated among the users. Second, the total transmit power is also optimized according to the solution of \textbf{P6}. However, no optimization is performed in the baseline schemes with respect to total transmit power. As a result, the time averaged throughput of the proposed scheme is better. 

\begin{figure}[h]
\begin{center}
\includegraphics[width=3.5in]{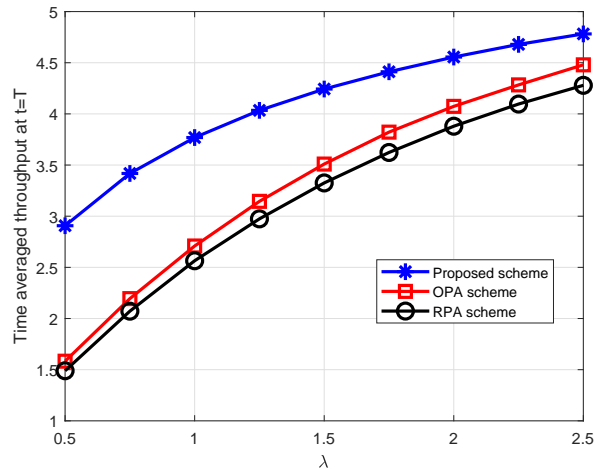}
\end{center}
\caption{Time averaged throughput with respect to $\lambda$ at $t=T$ for NOMA-WoR scenario.}
\end{figure}

Fig. 5 shows the dependence of the time averaged throughput at $t=T$ on $\lambda$. The throughput  increases with $\lambda$ for all the schemes. However, the performance gap between the proposed scheme and the baseline schemes becomes smaller. This is due to the following reason. As $\lambda$ increases the average harvested energy increases. This means the baseline schemes can transmit at relatively higher power to achieve a higher throughput. As a result the performance gap between the proposed scheme and baseline schemes shrinks. 

The performance comparisons of the proposed scheme and baseline schemes for OMA-WoR scenario are depicted in Fig. 6 and Fig. 7 with exponentially distributed channel gains and uniformly distributed channel gains, respectively. Again, it can be easily observed that the proposed scheme outperms the baseline scheme.
\begin{figure}[h]
\begin{center}
\includegraphics[width=3.5in]{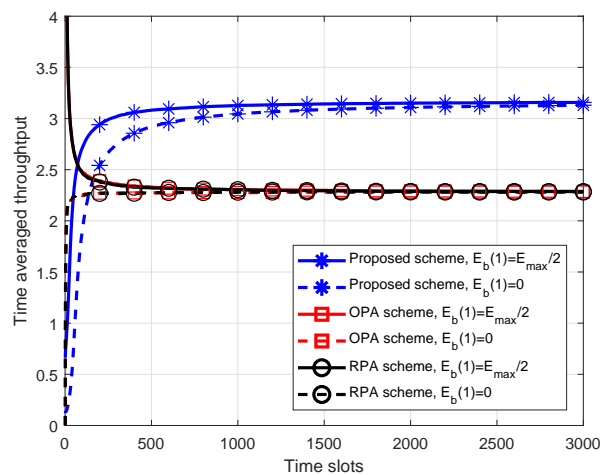}
\end{center}
\caption{Time averaged throughput with respect to time slots with when channel gains are exponentially distributed for OMA-WoR scenario.}
\end{figure}

\begin{figure}[h]
\begin{center}
\includegraphics[width=3.5in]{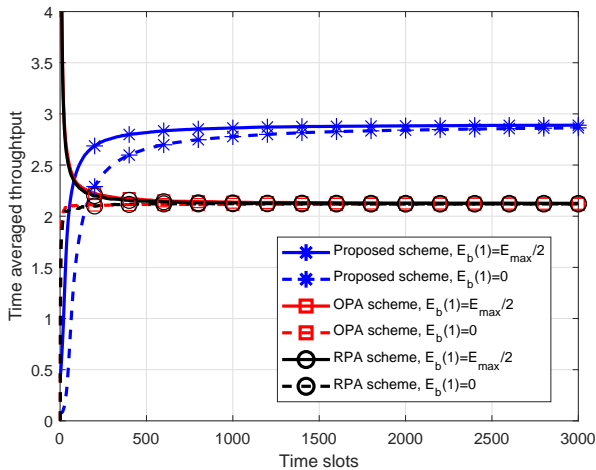}
\end{center}
\caption{Time averaged throughput with respect to time slots with when channel gains are uniformly distributed for OMA-WoR scenario.}
\end{figure}

In Fig. 8 we present the time averaged throughput for NOMA-WR, OMA-WR scenarios. Similar to the WoR scenario, the sum throughput converges to a definite value. Furthermore, as the rate constraints should be met for all the time slots, we observe that the minimum value of the time averaged throughput is $4$ since $R_{k,min}=1, \forall k \in \{1,\cdots, 4\}$. Another observation is that the time average sum throughput for the NOMA case is better than the OMA case. This performance improvement is a result of the variations in the channel gains of different users. 
\begin{figure}[h]
\begin{center}
\includegraphics[width=3.5in]{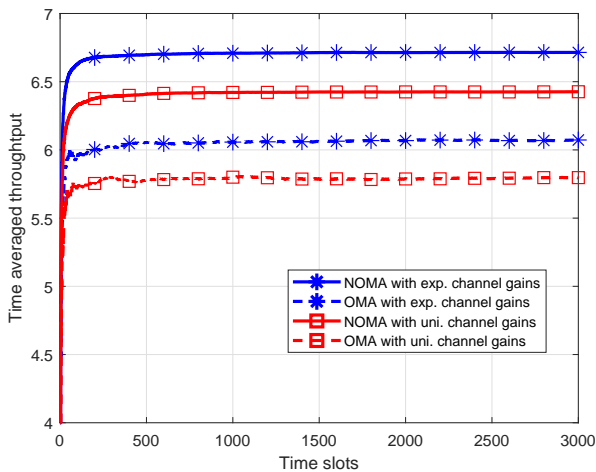}
\end{center}
\caption{Time averaged throughput comparison of the NOMA-WR and OMA-WR scenarios with respect to time slots with when channel gains are exponentially and uniformly distributed.}
\end{figure}
\section{Conclusions and Future Directions}
This paper developed online power allocation schemes for maximizing the time averaged sum rate for multiple downlink users. Two downlink transmission techniques are considered namely: NOMA and OMA. Specifically, we considered a scenario where the transmitter has energy harvesting and storage capabilities. Then, we have considered the scenario where individual users have a minimum rate requirement. The proposed schemes guarantee that the battery operational constraints are satisfied. The performance comparison between the optimal schemes and proposed schemes is carried out theoretically. It is shown that the proposed schemes have a bounded performance gap when compared with the optimal power allocation scheme. The simulation results demonstrate the effectiveness of the proposed schemes.

There are a few possible future directions that can be explored. Some of which we describe in the following. For example, as noted above that the proposed schemes for the case of individual rate requirements needs A1 and A2 to be valid. Therefore, an interesting future topic is to relax these assumptions and use machine learning techniques to maximize the long-term averaged sum rate. Another possible research contribution is to consider the possibility of hybrid downlink transmission where the base station employs NOMA for some users and OMA for remaining users. Another interesting research problem could be to investigate the multi-user cooperative scenario.     
\section*{Acknowledgement}
This work was supported by F.R.S.-FNRS under the EOS program (EOS project $30452698$) and by INNOVIRIS under the COPINE-IOT project.

\end{document}